\documentclass{llncs}
\usepackage{amssymb}
\usepackage{amsmath}
\usepackage{graphicx}
\usepackage{subfigure}
\newtheorem{procedure}{Procedure}
\newtheorem{algorithm}{Algorithm}
\usepackage{lineno}

\newcommand{\hide}[1]{}
\newcommand{\QED}{\hfill$\qed$}
\usepackage{color}

\usepackage{soul} %strikeout is \st{  }

\begin{document}
%\linenumbers
\title{The $p$-Center Problem in Tree Networks Revisited}%
\author{Aritra Banik \inst{1}
\and Binay Bhattacharya\inst{2}
\and Sandip Das\inst{1}
\and Tsunehiko Kameda \inst{2}
\and Zhao Song \inst{3}
}
\institute{Advanced Computing and Microelectronics Unit, Indian Stat. Inst., Kolkata, India %[sandipdas@isical.ac.in] 
\and School of Computing Science, Simon Fraser University, Burnaby, Canada %[\{binay,tiko\}@sfu.ca]
\and Department of Computer Science, University of Texas, Austin, U.S.A.
}

\maketitle

\begin{abstract}
We present two improved algorithms for weighted discrete $p$-center problem
for tree networks with $n$ vertices.
One of our proposed algorithms runs in $O(n \log n + p \log^2 n \log(n/p))$ time.
For all values of $p$,
our algorithm thus runs as fast as or faster than the most efficient $O(n\log^2 n)$ time
algorithm obtained by applying Cole's speed-up technique~\cite{cole1987}
to the algorithm due to Megiddo and Tamir~\cite{megiddo1983c},
which has remained unchallenged for nearly 30 years.

Our other algorithm,
which is more practical,
 runs in $O(n \log n + p^2 \log^2(n/p))$ time,
and when $p=O(\sqrt{n})$ it is faster than Megiddo and Tamir's $O(n \log^2n \log\log n)$
time algorithm~\cite{megiddo1983c}.
\end{abstract}

%%%%%%
\section{Introduction}\label{sec:introduction}
Deciding where to locate facilities  to minimize the communication or travel costs
is known as the {\em facility location problem}.
It has attracted much research interest since the publication of
the seminal paper on this topic by Hakimi~\cite{hakimi1964}.
For a good review of this subject, the reader is referred to \cite{hale2003}.
It can be applied to locate fire stations, distribution centers, etc.

In the {\em $p$-center problem},
$p$ centers are to be located in a network $G(V,E)$,
so that the maximum (weighted) distance from
any demand point to its nearest center is minimized.
The simplest version of the problem (V/V/p) allows centers to be
located only on vertices (V), and
restricts demand points to be vertices.
Other variations allow points on edges to be demand points
(V/E/p), or points on edges (E) to be centers (E/V/p), or both (E/E/p).
The vertices of a network could be weighted,
i.e., the vertex weights can be different,
or unweighted.
In this paper we refer to weighted E/V/p as the {\em weighted discrete} $p$-center problem ({\em WD$p$C}).
The $p$-center problem in a general network is NP-hard~\cite{kariv1979b}.
In this paper, we focus on  the tree networks,
on which there has been very little progress (for arbitrary $p$) since the mid-1980s.

\subsection{Previous work}
Megiddo~\cite{megiddo1983a} solved E/V/1 for the tree networks
in $O(n)$ time,
where $n$ is the number of vertices.
Megiddo and Tamir also studied this problem~\cite{megiddo1983c}.
Kariv and Hakimi \cite{kariv1979b} presented
an $O(m^p n^{2p-1} \log n/(p\!-\!1)!)$ time algorithm for WD$p$C in a general network,
where $m$ is the number of edges.
Tamir \cite{tamir1988} improved the above bound to $O(m^p n^p \log n\alpha'(n))$,
where $\alpha'(n)$ is the inverse of Ackerman's function.
Recently, Bhattacharya and Shi~\cite{bhattacharya2014b} improved it to
$O(m^pn^{p/2}2^{\log ^*n}\log n)$ for $p\geq 3$,
where $\log ^* n$ denotes the iterated logarithm of $n$.
A recent result on Klee's measure due to Chan~\cite{chan2013} implies that
this bound can be further improved to $O(m^pn^{p/2}\log n)$.

Frederickson \cite{frederickson1990,frederickson1991b} solved the unweighted
V/V/p, E/V/p and V/E/p problems in $O(n)$ time, independently of $p$.
For the weighted tree networks, linear time algorithms have been proposed
in the case where $p$ is a constant~\cite{benmoshe2006,shi2008}.
For arbitrary $p$, Kariv and Hakimi~\cite{kariv1979b} gave an exhaustive $O(n^2\log n)$ time
algorithm.
Megiddo's linear time feasibility test~\cite{megiddo1981} can be parameterized
to solve the problem in $O(n^2)$ time,
using the idea introduced in \cite{megiddo1981}.
Megiddo and Tamir~\cite{megiddo1983c} then provided
an $O(n\log^2 n\log\log n)$ time algorithm,
which can be made to run in $O(n\log^2n)$ time using the AKS or similar
$n \times O(\log n)$ sorting networks~\cite{ajtai1983,goodrich2014,seiferas2009},
together with Cole's improvement~\cite{cole1987}.
The $O(pn \log n)$ time algorithm due to Jeger and Kariv~\cite{jeger1985}
is faster than all others if $p =o(\log n)$.

The running time of the algorithm of Megiddo and Tamir~\cite{megiddo1983c}
is dominated by the time for computing the distance queries in their binary-search based algorithm.
Frederickson \cite{frederickson1990,frederickson1991b} used parametric search
to design optimal algorithms for the unweighted $p$-center problem in tree networks.
In parametric search,
one first designs an {$\alpha$-{\em feasibility} test to see if $p$ centers can be placed
in such a way that every vertex is within {\em cost} (=distance weighted by the weight of the vertex)
$\alpha$ from some center.
In general, a set of candidate values for $\alpha$ is explicitly or implicitly tested
as the algorithm progresses.
Eventually, the search will settle on the smallest $\alpha$ value, $\alpha^*$.
The ideas presented in \cite{frederickson1990,frederickson1991b} are for the unweighted case only,
and therefore cannot be extended easily to WD$p$C.
The question of whether an algorithm which runs faster than $O(n\log^2n)$ time
is possible for the tree networks has been open for a long time since.

To present our basic approach clearly,
we first solve WD$p$C for balanced binary tree networks.
We then generalize it to general (unbalanced) tree networks based on {\em spine tree decomposition}~\cite{benkoczi2004,benkoczi2003}.

%%%%%
\subsection{Our contributions:}
Our major contributions in this paper are
(i) an $O(p\log(n/p))$ time algorithm for testing $\alpha$-feasibility for an arbitrary $\alpha$,
with preprocessing that requires $O(n\log n)$ time,
(ii) a practical $O(n\log n + p^2\log^2(n/p))$ time WD$p$C algorithm,
which outperforms the $O(n\log^2 n\log\log n)$ time algorithm proposed in~\cite{megiddo1983c}
when $p=O(\sqrt{n})$,
and
iii)  an $O(n\log n+p\log^2 n\log(n/p))$ time WD$p$C algorithm based on AKS-like sorting
networks~\cite{ajtai1983,goodrich2014,seiferas2009},
which improves upon the currently best $O(n\log^2n)$ time algorithm~\cite{cole1987,megiddo1983c}.

The rest of the paper is organized as follows.
In Section~\ref{sec:preliminary} we first define the terms that are used throughout the paper.
We then give a rough sketch of our parametric search approach to solving WD$p$C on balanced tree networks.
We also propose our location policy that guides the placement of the centers.
Section~\ref{sec:preproc} describes preprocessing that we perform,
in particular, the computation of upper envelopes and
a preparation for fractional cascading.
We then present in Section~\ref{sec:feasibility} the details of the feasibility test
part of parametric search for balanced tree networks.
The optimization part of parametric search is discussed in detail in Section~\ref{sec:optimal}
for balanced tree networks.
At the end of the section,
we present our results for the general (unbalanced) tree networks.

%%%%%%
\section{Preliminaries}\label{sec:preliminary}

%%%%%%
\subsection{Definitions}\label{sec:def}
Let $T\!=\!(V,E)$ denote a tree network,
where each vertex $v\in V$ has weight $w(v)~(\geq 0)$
and each edge $e\in E$ has a non-negative length.
We write $x\in T$, if point $x$ lies anywhere in $T$, be it on an edge or at a vertex.
For $a,b\in T$, 
let $\pi(a, b)$ denote the unique path from $a$ to $b$,
and $d(a, b)$ its length.
If $a$ or $b$ is on an edge,
its prorated length is used.
If $T$ is a binary rooted with root vertex $r$,
for any vertex $v\in V$, 
the subtree rooted at $v$ is denoted by $T(v)$,
and the parent of $v~(\not=r)$ is denoted by $p(v)$.

For a non-leaf vertex $v\in V$,
let $v_l$ (resp. $v_r$) denote its left (resp. right) child vertex,
and define the left (resp. right) {\em branch} of $v$ by
$B(v_l)=T(v_l) \cup (v_l,v)$ (resp. $B(v_r)=T(v_r) \cup (v_r,v)$).
We thus have $T(v) = B(v_l) \cup B(v_r)$,
and the root of $B(v_l)$ (resp. $B(v_r)$) is $v$ with degree 1
in $B(v_l)$ (resp. $B(v_r)$).

Let $V' \subseteq V$ and $x\in T$.
We define the distance between a point $x$ and $V'$ by
$
d(x,V') \triangleq \min_{v\in V'}\{d(x,v)\}.
$
The {\em cost} of a vertex $v$ at point $x$ is given by $d(v,x)w(v)$.
We say that point $x\in T$ {\em $\alpha$-covers} $V'~(\subseteq V)$
if $\max_{v\in V'}\{d(x,v)w(v)\} \leq \alpha$.
If $\alpha$ is clear from the context, we may simply say that $x$ {\em covers} $V'$.
A problem instance is said to be {\em $\alpha$-feasible} if there exists $p$ centers
such that every vertex is $\alpha$-covered by at least one of the centers.
Those $p$ centers are said to form a {\em p-center}~\cite{kariv1979b}. 
For a vertex $v\in V$ and points $x\in T\!\setminus\! T(v)$,
we define the {\em upper envelope}
\begin{equation}\label{eqn:uppEnv0}
E_v(x) = \max_{u\in T(v)}\{d(x,u)w(u)\}.
\end{equation}
If $E_v(x) =d(u,x)w(u)=\alpha$, 
then vertex $u$ is said to be an $\alpha$-{\em critical vertex} in $T(v)$ with respect to $x\in T\!\setminus\! T(v)$,
and is denoted by $u={\it cv}(x,T(v))$.  
If $\alpha$ is clear from the context,
we may call it just a critical vertex

%%%%%%
\subsection{Spine tree decomposition and upper envelopes}\label{sec:std}
We give a brief review of {\em spine tree decomposition}~\cite{benkoczi2004,benkoczi2003}.
The materials in this subsection is not needed until Sec.~\ref{sec:general}.
We can assume that given $T$ is a binary tree;
otherwise we can introduce $O(n)$ vertices of 0 weight and $O(n)$ edges of
0 length to make it binary.
Thus each vertex has degree at most 3.
Let $r$ be the root of $T$,
which can be chosen arbitrarily.
Traverse $T$, starting on an edge incident to $r$.
At each vertex visited,
move to the branch that contains the largest number of leaf vertices,
breaking a tie arbitrarily.
When a leaf vertex, $u$, is reached, the path $\pi(v,u)$ is generated,
and it is called the {\em top spine}, denoted by $\sigma_1$.
We then repeat a similar traversal from each vertex on the generated spine,
to generate other spines, until every vertex of $T$ belongs to some spine.

Let ${\it STD}(T)$ denote the tree constructed by the spine tree decomposition of tree $T$,
together with the {\em search tree} $\tau_{\sigma_l}$ for each {\em spine} $\sigma_l$,
whose root is denoted by $\rho_l$~\cite{benkoczi2004,benkoczi2003}.
Fig.~\ref{fig:uppEnvTree}  illustrates a typical structure of spine $\sigma_l$ and
its search tree $\tau_{\sigma_l}$.
The horizontal line represents spine $\sigma_l$,
and we name the vertices on it  $v_1, v_2, \ldots$ from left to right. 
\begin{figure}[ht]
\centering
\includegraphics[height=2.5cm]{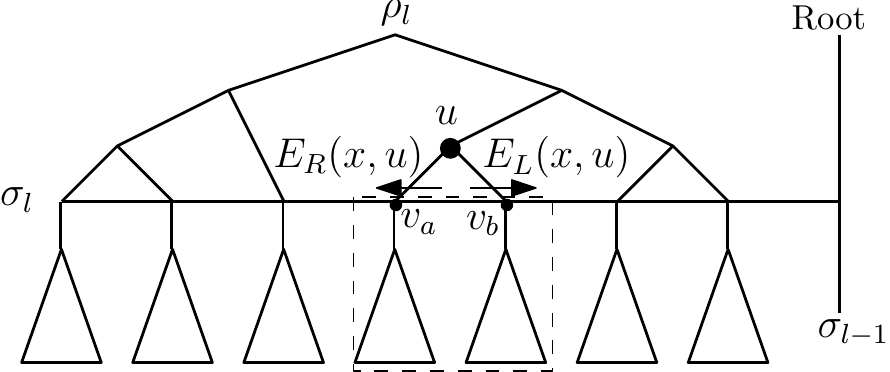}
\caption{Search tree $\tau_{\sigma_l}$ for spine $\sigma_l$.
$v_a=v_L(u)$ and $v_b=v_R(u)$. 
}
\label{fig:uppEnvTree}
\end{figure}
The triangles represent subtrees hanging from $\sigma_l$.
If a hanging subtree $t$ is connected to vertex $v_i\in \sigma_l$,
then we call the subgraph consisting of $t$, $v_i$, and the edge connecting them a {\em branch}
of $\sigma_l$ and denote it by $B_i$.
Since we assume that the vertices of $T$ have degree at most 3,
there is at most one branch hanging from any vertex on the spine.
 
For a node\footnote{A `node' is more general than a vertex of $T$.
A vertex is also a node, because it belongs to $\tau_{\sigma_l}$,
but not every node is a vertex.}
$u$ in $\tau_{\sigma_l}$,
let $v_L(u)$ (resp. $v_R(u)$) denotes the leftmost
(resp. rightmost)\footnote{Right (resp. left) means towards (resp. away from)
the parent spine $\sigma_{l-1}$ of $\sigma_l$.}
 vertex on $\sigma_l$ that belongs to the subtree $\tau_{\sigma_l}(u)$.
We introduce upper envelope $E_L(x,u)$ (resp. $E_R(x,u)$)
for the costs
of the vertices in the branches of $\sigma_l$ that belong to $\tau_{\sigma_l}(u)$,
for point $x$ that lies to the right (resp. left) of vertex $v_R(u)$ (resp. $v_L(u)$). 
See Fig.~\ref{fig:uppEnvTree}.
Since $E_L(x,u)$ and $E_R(x,u)$ are upper envelopes of linear functions,
they are piecewise linear.
For each node $u$ of ${\it STD}(T)$ we compute $E_L(x,u)$ and $E_R(x,u)$,
and store them at $u$ as sequences of bending points
(their $x$ and $y$ coordinates).
These upper envelopes can be computed in $O(n\log n)$ time
by the following lemma.
\begin{lemma}\label{lem:std}{\rm \cite{benkoczi2004,benkoczi2003}}
The path from any leaf to the root of ${\it STD}(T)$ has $O(\log n)$ nodes on it.
\QED
\end{lemma}

%%%%%%
\subsection{Our approach}\label{sec:approach}
Except in the last subsection of the paper,
we assume that the given tree $T$ is balanced with respect to its root $r$,
so that its height is $O(\log n)$.
If not, we can use spine tree decomposition that transforms $T$ in linear time to a structure
that has most of the properties of a balanced binary tree.
Working on a balanced binary tree network also helps us to explain the essence of our approach,
without getting bogged down in details. 
Our algorithms consist of a lower part and an upper part.
In the lower part, we test $\alpha$-feasibility for a given cost $\alpha$,
and in the upper part we carry out Megiddo's {\em parametric search}~\cite{megiddo1979}.
To perform a feasibility test,
we first identify the $\alpha$-{\em peripheral centers},
below which no center needs be placed.
Once all the $q~(< p)$ $\alpha$-peripheral centers are identified,
we place $p - q$ additional centers to $\alpha$-cover the vertices
that are not covered by the $\alpha$-peripheral centers.
If no more than $p$ centers are used to $\alpha$-cover the entire tree $T$,
then the $\alpha$-feasibility test is successful.
Theorem~\ref{thm:feasibility} shows that, using fractional cascading, $\alpha$-feasibility can be tested
in $O(p\log(n/p))$ time
after preprocessing, which takes $O(n\log n)$ time.

The second part of parametric search
finds the smallest $\alpha$ value, $\alpha^*$.
We work on $T$ bottom-up, doing essentially the same thing as in the first part.
Whenever $\alpha$ is used in the first part,
we need to invoke an $\alpha$-feasibility test~\cite{megiddo1979}.
At each level of $T$,
we need to invoke $\alpha$-feasibility tests $O(l)$ times at level $l$.
Therefore the total number of invocations is $O(\log^2 n)$,
and the total time is $O(p\log^2n\log(n/p))$ after preprocessing,
yielding one of our main results stated in Theorem~\ref{thm:main}.

%%%%
\subsection{Center location policy}
Suppose that we want to place a center $c_i$ in a tree network $T$ to $\alpha$-cover
a subset $V_i$ of vertices that are connected.
We propose the following location policy.

\medskip\noindent
{\bf Root-centric policy:}
{\em Place $c_i$ at the point that $\alpha$-covers all the vertices in $V_i$ and is closest to root $r$ of $T$.}

It is easy to prove the following lemmas.
\begin{lemma}\label{lem:location}
If a set of $p$ centers $\alpha$-covers all the vertices in $V$,
then there is a partition of vertex set $\{V_i\mid i = 1, \ldots, p\}$,
where each $V_i$ is the vertex set of a connected part of $T$,
such that the root-centric location policy locates each center $c_i$ that $\alpha$-covers $V_i$.
\QED
\end{lemma}
\hide{%%
\begin{proof}
Consider a set $C=\{c_i\mid i = 1, \ldots, p\}$ of centers that together $\alpha$-covers $V$,
but some centers violate the root-centric location policy.
For each $c_i \in C$, 
let $V_i$ be the set of vertices that are closer to $c_i$ than to any other center.
If there is a tie, then the tie should be broken arbitrarily.
It is clear that $V_i$ is $\alpha$-covered by $c_i$.
We now move $c_i$ towards the root $r$ as far as possible without affecting its coverage
of $V_i$.
The resulting centers $\{c'_i\mid i = 1, \ldots, p\}$ satisfy the root-centric location policy.
\end{proof}
}%end

\begin{lemma}\label{lem:Rcentric}
Let  $\{c_i\mid i = 1, \ldots, p\}$ be $p$ centers obeying the root-centric policy
that together $\alpha$-cover $V$.
For each center $c_i$, find a vertex $v\in V_i$ with maximum cost $d(v,c_i) w(v)$
that is the farthest from the root,
and name it $g_i$.
Then it satisfies $c_i \in \pi(g_i,r)$.
\end{lemma}
\begin{proof}
If $c_i \notin \pi(g_i,r)$, then $c_i$ could move closer to $r$,
a contradiction.
\QED
\end{proof}

%%%%%%
\section{Preprocessing}\label{sec:preproc}

%%%%%%%
\subsection{Upper envelopes}
According to our definition of upper envelope $E_v(x)$ for subtree $T(v)$ (see (\ref{eqn:uppEnv0})),
if $v$ is a leaf vertex, we have
\begin{equation}\label{eqn:uppEnv1}
E_v(x) = d(x,v)w(v),
\end{equation}
for any $x \in T$.
Let $v_l$ (resp. $v_r$) be the left (resp. right) child vertex of a non-leaf vertex $v\in V$.
Then for any $x\in T\setminus T(v)$, we have
\begin{equation}\label{eqn:uppEnv2}
E_v(x) = \max\{E_{v_l}(x),E_{v_r}(x),d(x,v)w(v)\}.
\end{equation}
Function $E_v(x)$ is piecewise linear in $x\in \pi(v,r)$ and can be represented by a sequence of
bending points.
In the sequence representing $E_v(x)$,
in addition to the values of $E_v(x)$ at the bending points,
we insert the values of $E_v(x)$ evaluated at all the $O(\log n)$ vertices on $\pi(v,r)$.\footnote{We mix
those values among the bending points,
so that we know on which edges the bending points lie.}
\begin{lemma}\label{lem:envelope}
If $T$ is balanced, then 
$\{E_v(x) \mid v\in V, x\in \pi(v,r)\}$ can be computed bottom-up in $O(n\log n)$ time
and $O(n \log n)$ space.
\QED
\end{lemma}

In the rest of this paper we assume that the given tree $T$ is a balanced binary tree.
If not we can use spine tree decomposition~\cite{benkoczi2004,benkoczi2003,bhattacharya2012b},
which shares many useful properties of a balanced tree.

%%%%%%%
\subsection{Fractional cascading}\label{sec:fractional}

From now on we assume that we have the bending points of
$\{E_v(x)\mid v \in V,x\in \pi(v,r)\}$ at our disposal.
The second task of preprocessing is to merge the bending points of $\{E_v(x)\mid v \in V, x\in \pi(v,r)\}$
to prepare for fractional cascading~\cite{chazelle1986a}.
Again we do this bottom up, 
merge-sorting the two sequences of bending points into one at each vertex.
Since each vertex causes at most $O(\log n)$ bending points in $\{E_v(x)\mid v \in V, x\in \pi(v,r)\}$,
the total number of bending points is $O(n \log n)$.

%%%%%%
\section{$\alpha$-Feasibility}\label{sec:feasibility}

%%%%
\subsection{Peripheral centers}\label{sec:Pctrs}
As a result of preprocessing,
we have the upper envelopes $\{E_v(x)\mid v\in V, x\in \pi(v,r)\}$.
To find the peripheral centers, $\alpha$-peripheral 
we carry out {\em truncated} pre-order DFS (depth-first-search),
looking for the vertex-point pairs $(v,x)$ satisfying $E_v(x) = \alpha$,
which means $v$ is an $\alpha$-critical vertex in $T(v)$ with respect to $x$.

%%%%%
\begin{procedure}{\tt Find-Peripheral-Centers}\label{proc:findPC}$(\alpha)$

~\noindent
Perform pre-order DFS, modified as follows,
where $v$ is the vertex being visited.
\begin{enumerate}
\item
If $\exists x\in (v,p(v))$ such that $E_v(x)= \alpha$,
return $x$ as an $\alpha$-peripheral center,\footnote{We assume that the trivial case,
where one center at root $r$ $\alpha$-covers the entire tree, is dealt with specially,
which is straightforward. 
}
and backtrack.
\item
If $p\!+\!1$ $\alpha$-peripheral centers have been found, then return {\tt Infeasible} and stop.
\QED
\end{enumerate}
\end{procedure}
To carry out Step~1 efficiently,
we perform binary search with key $\alpha$ in the merged sequence
of bending points (of the upper envelopes) stored at the root $r$,
and follow the relevant pointers based on fractional cascading.

\begin{lemma}\label{lem:pathLength}
Procedure~{\tt Find-Peripheral-Centers}$(\alpha)$ visits 
$O(p\log(n/p))$ vertices.
\end{lemma}
\begin{proof}
The number of vertices that Procedure~{\tt Find-Peripheral-Centers}$(\alpha)$
visits is the largest when the $\alpha$-peripheral centers are as low as possible
and they separate from each other as high as possible.
This extreme case is illustrated in Fig.~\ref{fig:pathLength},
where $p= 2^{k}-1$ for some integer $k$.
\begin{figure}[h]
\centering
\includegraphics[height=2.6cm]{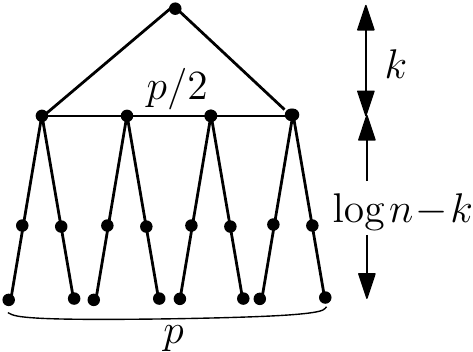}
\caption{Illustration for the proof of Lemma~\ref{lem:pathLength}.}
\label{fig:pathLength}
\end{figure}
The total number of edges that are traversed is given by
\[
O(p(\log n -k) +p) = O(p(\log n - \log p) +p) =O(p\log (n/p)),
\]
where the second term, $p$, is an upper bound on the number of vertices at depth $k$ or shallower.
\QED
\end{proof}

\begin{lemma}\label{lem:findPC}
If $\{E_v(x)\mid v\in V\}$ 
are available,
all the $\alpha$-peripheral centers can be found in $O(p\log(n/p))$ time.
\end{lemma}
\begin{proof}
If fractional cascading is used in Step~1 of Procedure~\ref{proc:findPC},
it runs in amortized constant time per vertex.
The rest follows from Lemma~\ref{lem:pathLength}.
\QED
\end{proof}

%%%%%%
\subsection{$\alpha$-Feasibility test}\label{sec:test}
Given an $\alpha$ value,
suppose that we have found $q ~(<p)$ $\alpha$-peripheral centers,
following the root-centric location policy.
We replace each $\alpha$-peripheral center by a {\em dummy vertex},
and define the {\em trimmed tree} $T'_{\alpha}=(V'_{\alpha},E'_{\alpha})$.
Its vertex set $V'_{\alpha}$ consists of two types of vertices:
the {\em first type} is a vertex that lies on the path between a dummy vertex and root $r$, inclusive.
If any such vertex has only one child vertex among them,
then the other child vertex of $T$ (called a vertex of the {\em second type}) is kept in $T'$
to represent the $\alpha$-critical vertex in the subtree of $T$ rooted at that vertex.
In what follows, we use $T'$ instead of $T'_{\alpha}$ for simplicity,
since the implied $\alpha$ will be clear from the context.
It is easy to see that tree $T'$ contains $O(q\log n)$ vertices.
Without loss of generality,
we consider each vertex of the second type as the right child of its parent.

Let $u$ be a vertex of the second type.
Then we must have visited $u$ during the execution of {\tt Find-Peripheral-Centers}$(\alpha)$,
and no $\alpha$-peripheral center was placed in subtree $T(u)$.
At the time of this visit,
we identified the $\alpha$-critical vertex in $T(u)$,
which implies that we can store this $\alpha$-critical vertex at $u$ as a by-product of
{\tt Find-Peripheral-Centers}$(\alpha)$ at no extra cost.

Later, we will be introducing more centers, in addition to $\alpha$-peripheral centers,
working on the trimmed tree $T'$ bottom up.
For each vertex in $T'$,
its subtrees can be one of the following types:
\begin{enumerate}
\item[-]
$\ominus$-subtree: The centers in it, if any, do not $\alpha$-cover all the vertices in the subtree.
\item[-]
$\oplus$-subtree: The centers in it $\alpha$-cover all the vertices in the subtree, and possibly outside it.
\end{enumerate}

If $T'(v)$ is a $\oplus$-subtree,
let $\delta_+^{\alpha}(v)$ denote the distance from $v$ to the highest center in $T'(v)$
at or below $v$.
See the leftmost figure of Fig.~\ref{fig:covering1},
where $v_l$ (resp. $v_r$) is the left (resp. right) child vertex of $v$,
and $c_l$ (resp. $c_r$) is the highest center placed in $T'(v_l)$ (resp. $T'(v_r)$).
\begin{figure}[ht]
\centering
\includegraphics[height=2.6cm]{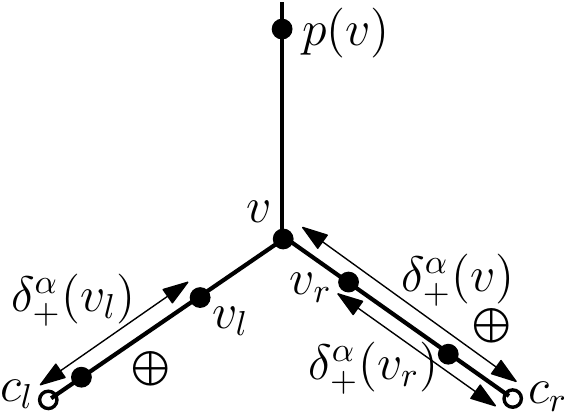}
\hspace{8mm}
\includegraphics[height=2.6cm]{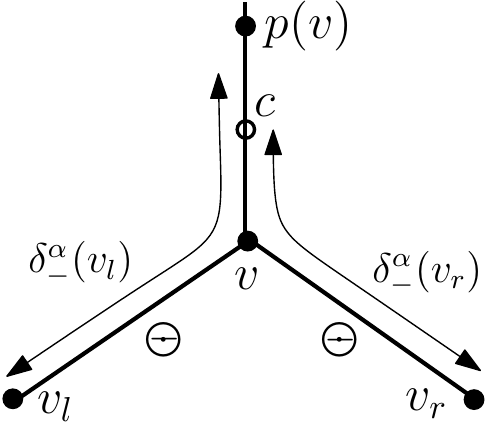}
\hspace{8mm}
\includegraphics[height=2.6cm]{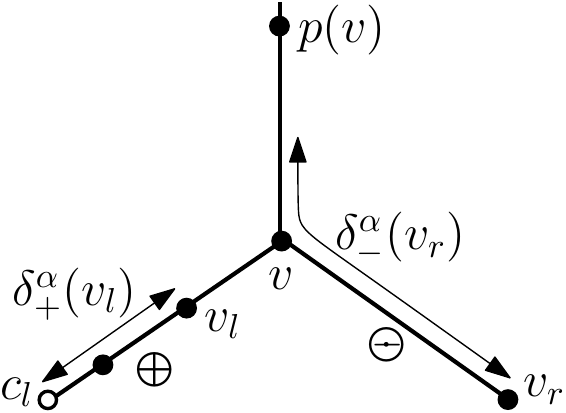}
\caption{(Left) $c_l\in B(v_l)$ and $c_r\in B(v_r)$; 
(Middle) A center is needed within $\delta_-^{\alpha}(u)$ from $u$;
(Right) $c_l\in B(v_l)$.
}
\label{fig:covering1}
\end{figure}
If $T'(v)$ is a $\ominus$-subtree,
on the other hand,
let $\delta_-^{\alpha}(v)$ denote the minimum distance from $v$ to a point above $T'(v)$
within which a center must be placed to $\alpha$-cover the uncovered vertices in $T'(v)$.
See the middle figure in Fig.~\ref{fig:covering1}.

Let us discuss how to process the trimmed tree $T'$,
to introduce additional centers closer to the root in order to $\alpha$-cover more vertices.
We perform post-order DFS on $T'$,
always visiting the left child of a vertex first.
Assume that we explored $T'(v_l)$ first and then $T'(v_r)$,
and we are just back to $v$,
and that $\delta_-^{\alpha}(v_l)$ or $\delta_+^{\alpha}(v_l)$ (resp. $\delta_-^{\alpha}(v_r)$ or $\delta_+^{\alpha}(v_r)$)
are available at vertex $v_l$ (resp. $v_r$).
For each dummy leaf vertex $v$ of $T'$,
we have $\delta_+^{\alpha}(v)=0$.
At each vertex $v$ visited, we have one of the following three cases.

(a) {\bf [Both are $\oplus$-subtrees]}
In the leftmost figure of Fig.~\ref{fig:covering1},
$c_l$ (resp. $c_r$) is the highest center in $T'(v_l)$ (resp. $T'(v_r)$).
We compute 
\begin{equation}\label{eqn:updateDelta+}
\delta=\min \{\delta_+^{\alpha}(v_l)+ d(v,v_l),\delta_+^{\alpha}(v_r)+ d(v,v_r)\},
\end{equation}
which is the distance from $v$ to the nearest center in $T'(v)$.
If $\delta\cdot w(v) \leq \alpha$,
then $v$ is $\alpha$-covered by $c_l$ or $c_r$. 
Otherwise (i.e., even the center in $T'(v)$ that is nearer to $v$ cannot $\alpha$-cover $v$)
$v$ must be covered by a center placed above $v$,
and $T'(v)~(=\{v\})$ now becomes a $\ominus$-subtree of $p(v)$.

(b) {\bf [Both are $\ominus$-subtrees]} See the middle figure of  Fig.~\ref{fig:covering1}.
If $\delta_-^{\alpha}(v_l) <d(v,v_l)$,
for example,
we need to place a center $c_l$ on the edge $(v,v_l)$, 
and $T'(v)$ now becomes a $\oplus$-subtree,
provided $v$ is $\alpha$-covered by $c_l$.
If both $c_l$ and $c_r$ are placed this way,
we set $\delta_+^{\alpha}(v)= \min\{d(c_l,v), d(c_r,v)\}$,
provided one of them $\alpha$-covers $v$.
If no center needs to be placed on $(v,v_l)$ or $(v,v_r)$,
then we compute
\begin{equation}\label{eqn:newDelta-}
\delta=\min\{\delta_-^{\alpha}(v_l) -d(v,v_l), \delta_-^{\alpha}(v_r)-d(v,v_r)\}.
\end{equation}
We need a center within $\min\{\delta,\alpha/w(v)\}$ above $v$.
These are some of the typical cases,
which illustrate kinds of necessary operations.
Procedure~{\tt Merge}$(v; \alpha, T)$, given below,
deals with the other cases as well, not mentioned here,
exhaustively.

(c) {\bf [One is a $\ominus$-subtree and the other is a $\oplus$-subtree]}
We assume without loss of generality that the left (resp. right) subtree is
a $\oplus$-subtree (resp. $\ominus$-subtree), as shown in the rightmost figure of  Fig.~\ref{fig:covering1},
and $c_l$ is the highest center in $T'(v_l)$.
As in Case (b),
we first test if $\delta_-^{\alpha}(v_r) < d(v,v_r)$,
and if so place a center $c_r$ on edge $(v,v_r)$.
Then we have case (a).
Otherwise,
we need to test if $c_l$ $\alpha$-covers the uncovered vertices in $T'(v_r)$
as well as $v$.
If not, they must be covered by a new center above $v$.

We now present a formal procedure that deals with all possible cases.
We will use it for $T= T'$.
%%%%%
\begin{procedure} {\tt Merge}$(v; \alpha, T)$\label{proc:merge}

\smallskip\noindent
{\bf Case (a):}
{\rm [$T(v_l)\!=\!\oplus, T(v_r)\!=\!\oplus$]}
Compute $\delta$ using (\ref{eqn:updateDelta+}).
If $\delta\cdot w(v) \leq \alpha$,
then set $\delta_+^{\alpha}(v)=\delta$. 
Otherwise, make $T(v)$ a $\ominus$-subtree of $p(v)$ with $\delta_-^{\alpha}(v)=\alpha/w(v)$.

\smallskip\noindent
{\bf Case (b):}
{\rm [$T(v_l)\!=\!\ominus, T(v_r)\!=\!\ominus$]}
If $\delta_-^{\alpha}(v_l) <d(v,v_l)$
(resp. $\delta_-^{\alpha}(v_r) < d(v,v_r)$),
place a center $c_l$ (resp. $c_r$)
on the edge $(v,v_l)$, (resp. $(v,v_r)$) at distance $\delta_-^{\alpha}(v_l)$ from $v_l$
(resp.  $\delta_-^{\alpha}(v_r) $ from $v_r$).
If $c_l$ and/or $c_r$ $\alpha$-covers $v$,
then make $T(v)$ a $\oplus$-subtree of $p(v)$ with
$\delta_+^{\alpha}(v)= \min\{d(c_l,v),$ $d(c_r,v)\}$,
where $d(c_l,v)=0$ (resp. $d(c_r,v)=0$) if
$c_l$ (resp. $c_r$) is not introduced.
If neither of them covers $v$,
then make $T(v)$ a $\ominus$-subtree of $p(v)$ with $\delta_-^{\alpha}(v)=\alpha/w(v)$.
If neither $c_l$ nor $c_r$ is introduced,
then compute $\delta$ using (\ref{eqn:newDelta-})
and make $T(v)$ a $\ominus$-subtree of $p(v)$ with $\delta_-^{\alpha}(v)=\min\{\delta,\alpha/w(v)\}$.

\smallskip\noindent
{\bf Case (c):}
{\rm [$T(v_l)\!=\!\oplus,T(v_r)\!=\!\ominus$]}\footnote{The case {\rm [$T(v_l)\!=\!\ominus,T(v_r)\!=\!\oplus$]}
 is symmetric.}
If $\delta_-^{\alpha}(v_r) < d(v,v_r)$,
then place a center $c_r$ on edge $(v,v_r)$ at distance $\delta_-^{\alpha}(v_r) $ from $v_r$,
set
$\delta_+^{\alpha}(c_r) = d(v,c_r)= d(v,v_r) - \delta_-^{\alpha}(v_r),$
and go to Case (a).
Otherwise,
\begin{enumerate}
\item[(i)]
If $c_l$ covers $v$ (i.e.,
$\{\delta_+^{\alpha}(v_l) + d(v_l,v))\} w(v) \leq \alpha$),
and $c_l$ also covers $T(v_r)$ (i.e.,
$\delta_+^{\alpha}(v_l) +d(v_l,v_r)\leq \delta_-^{\alpha}(v_r)$),
then  let $\delta_+^{\alpha}(v)=\delta_+^{\alpha}(v_l) + d(v_l,v)$.
\item[(ii)]
In all the remaining cases, set
$\delta_-^{\alpha}(v)=\min\{\delta_-^{\alpha}(v_r) -d(v,v_r),\alpha/w(v)\}$.
\QED
\end{enumerate}
\end{procedure}

It is easy to show that
\begin{lemma}\label{lem:merge}
After preprocessing,
{\tt Merge-I}$(v; \alpha, T)$ runs in constant time.
\QED
\end{lemma}

We now formally state our algorithm for testing $\alpha$-feasibility.
\begin{algorithm}{\tt Feasibility-Test}\label{alg:feasibility}$(\alpha,T)$
\begin{enumerate}
\item
Call {\tt Find-Peripheral-Centers}$(\alpha)$.
\item
Construct the trimmed tree $T'$, 
consisting of the vertices of the first type and those of the second type
and the edges connecting them.
For each vertex $u$ of the second type,
compute the $\alpha$-critical vertex for $T'(u)$.
\item
Perform a post-order depth-first traversal on $T'$,
invoking {\tt Merge}$(v; \alpha, T')$ on each vertex $v$ visited.
\item
If a set of no more than $p$ centers covering $T$ has been found,
then return {\tt Feasible} and stop.
If the $p$ centers found so far do not totally cover $T$,
then return {\tt Infeasible} and stop.
\QED
\end{enumerate}
\end{algorithm}

\begin{theorem}\label{thm:feasibility}
For a balanced tree network, 
{\tt Feasibility-Test}$(\alpha,T)$ runs in $O(p\log (n/p))$ time,
excluding the preprocessing time.
\end{theorem}
\begin{proof}
Step~1 runs in $O(p\log(n/p))$ time by Lemma~\ref{lem:findPC}.
Step~2 can be carried out at the same time as Step~1 in $O(p\log(n/p))$ time.
Step~3 also runs in $O(p \log (n/p))$ time  by Lemma~\ref{lem:merge}.
Lastly, Step~4 takes constant time.
\QED
\end{proof}

%%%%%%
\section{Optimization}\label{sec:optimal}
We will employ Megiddo's parametric search~\cite{megiddo1979},
using the $\alpha$-feasibility test we developed in Sec.~\ref{sec:test}.
We maintain a lower bound $\underline{\alpha}$ and an upper bound $\overline{\alpha}$ on $\alpha^*$,
where $\underline{\alpha}< \alpha^* \leq\overline{\alpha}$.
Eventually we will end up with $\alpha^* =\overline{\alpha}$.
If we succeed (resp. fail) in an $\alpha$-feasibility test, then it means that $\alpha\geq \alpha^*$
(resp. $\alpha< \alpha^*$), so we update $\overline{\alpha}$ (resp, $\underline{\alpha}$)
to $\alpha$.

%%%%%%
\subsection{Balanced tree networks}\label{sec:practical}
Based on Theorem~\ref{thm:feasibility},
the main theorem in~\cite{megiddo1979} implies
\begin{theorem}\label{thm:practical}
WD$p$C for the balanced tree networks with $n$ vertices can be solved in 
$O(n\log n + p^2\log^2(n/p))$ time.
\QED
\end{theorem}

We propose another algorithm which performs better than the first algorithm
referred to in the above theorem for some range of values of $p$.
For this algorithm we will show later that we need to test feasibility $O(\log^2 n)$ times.
This fact, together with Theorem~\ref{thm:feasibility}, leads to the following theorem.
\begin{theorem}\label{thm:main}
WD$p$C for the balanced tree networks with $n$ vertices can also be solved in 
$O(n \log n+p\log^2 n\log (n/p))$ time.
\QED
\end{theorem}

In the rest of this subsection we prove Theorem~\ref{thm:main}.
Let $l=1,2, \ldots,k$ be the {\em levels} of $T$ from top to bottom,
where the root $r$ is at level 1 and  the leaves are at level $k=\lceil \log n\rceil =O(\log n)$.
At each vertex, we need to perform a few feasibility tests.
Since there are $2^{l-1}$ vertices at level $l$ of $T$,
using prune and search, 
we can know the results of the feasibility tests at all the vertices of level $l$
after actually performing only $O(\log (2^{l-1}))=O(l)$ feasibility tests.
The total for all levels is thus $O(\sum_{l=1}^{\log n} l) =O(\log^2 n)$,
as claimed above.

It is easy to prove the following lemma.
\begin{lemma}\label{lem:eqCost}
Let $v_a, v_b\in V$.
\begin{enumerate}
\item[(a)]
{\rm \cite{kariv1979b}}
Vertices $v_a$ and $v_b$ have the equal cost
\begin{equation}\label{eqn:equalcost1}
\alpha(v_a, v_b) 
= \frac{d(v_a, v_b)w(v_a)w(v_b)}{w(v_a)+w(v_b)}
\end{equation}
at a point $c(v_a, v_b)\in \pi(v_a, v_b)$.
\item[(b)]
Let $v_a, v_b \in T(v)$, and
suppose that $w(v_a)\not= w(v_b)$, and let $w(v_a)< w(v_b)$ without loss of generality.
If $d(v_a, v)w(v_a) \geq d(v_b, v)w(v_b)$ holds,
then vertices $v_a$ and $v_b$ have the equal cost
\begin{equation}
\alpha'(v_a, v_b) 
= \frac{\{d(v_a, v)- d(v_b, v)\}w(v_a)w(v_b)}{w(v_b)-w(v_a)}, \label{eqn:equalcost2}
\end{equation}
at a point $c'(v_a, v_b)\in \pi(v,r)$. 
If $d(v_a, v)w(v_a) < d(v_b, v)w(v_b)$,
then vertex $v_b$ has a higher cost than $v_a$ at all points on $\pi(v,r)$.\footnote{In this case,
the equal cost point lies on $\pi(v, v_b)$.}
\QED
\end{enumerate}
\end{lemma}

If we let $v_b=v$ in Case (b) in the above lemma,
$v_a$ and $v$ have the equal cost
\begin{equation}
\alpha'(v_a, v) 
= \frac{d(v_a, v)w(v_a)w(v)}{w(v)-w(v_a)}, \label{eqn:equalcost4}
\end{equation}
at a point $c'(v_a, v_b)\in \pi(v,r)$.

We now need to modify the definition of the critical vertex given in Sec.~\ref{sec:def}.
With respect to $x\in T\setminus T(v)$,
we are interested in the vertex $u\in T(v)$, 
such that $\alpha(x, u)$ is maximum,
We call such a $u$ the {\em critical vertex} with respect to $x$ and denote it by $\gamma_v$.
The main difference of the optimization part from the feasibility test part
is that we cannot find the exact locations of the centers until the very end.
However, making use of critical vertices, it is possible to identify
the component of $T$ that is to be $\alpha^*$-covered by each new center.
So, we will isolate/detach them one by one from $T$,
and repeat the process.

Let $v_l$ and $v_r$ be the two child vertices of a vertex $v$ at level $l$.
When we visit $v$, moving up $T$,
we need to either isolate a subtree to be covered by a center
that lies below $v$,
or determine the critical vertex in $T(v)$ to be carried higher.
Whenever the result of an $\alpha$-feasibility test shows that $\alpha\geq \alpha^*$,
we update $\overline{\alpha}$ and assume that $\overline{\alpha}> \alpha^*$ holds,
and introduce a new center (without an exact location),
as necessary.
This assumption will be justified if $\overline{\alpha}$ is updated later.
If $\overline{\alpha}$ is never updated thereafter,\footnote{$\underline{\alpha}$
may be updated.}
it implies that $\overline{\alpha}=\alpha^*$.
See Lemma~\ref{lem:finalAlpha}.

Based on (\ref{eqn:equalcost1}), if 
\begin{equation}\label{eqn:test10}
\alpha(v,\gamma_{v_l})  \geq  \alpha^*
\mbox{~(resp. ~}
\alpha(v,\gamma_{v_r}) \geq \alpha^*\mbox{)},
\end{equation}
we assume that $\alpha(v,\gamma_{v_l})  >  \alpha^*\mbox{~(resp. ~}
\alpha(v,\gamma_{v_r}) > \alpha^*\mbox{)}$, 
and cut the edge $(v,v_l)$ (resp. $(v,v_r)$) to detach a new component below $v$
to be covered by the new center placed in it.\footnote{Note that if
$\alpha(v,\gamma_{v_l}) =  \alpha^*$, for example,
we cannot isolate a component.}
We need not know the exact position of the new center.
If two new centers are introduced this way,
vertex $v$ must be $\alpha^*$-covered by a center placed above $v$,
and $v$ becomes a (tentative) critical vertex for $T(v)$ with respect to $x$ above $v$.
If only one of the inequalities in (\ref{eqn:test10}) holds and only $(v,v_l)$ (resp. $(v,v_r)$) is cut,
then either $v$ or $\gamma_{v_r}$ (resp. $\gamma_{v_l}$) becomes a critical vertex for $T(v)$,
based on the outcome of $\alpha'(v_a, v)$-feasibility test.
See (\ref{eqn:equalcost4}).

Consider the remaining case,
 where neither inequality in (\ref{eqn:test10}) holds.
We need to determine a critical vertex in $T(v)$ with respect to $x$ above $v$.
\begin{figure}[ht]
\centering
\includegraphics[height=2cm]{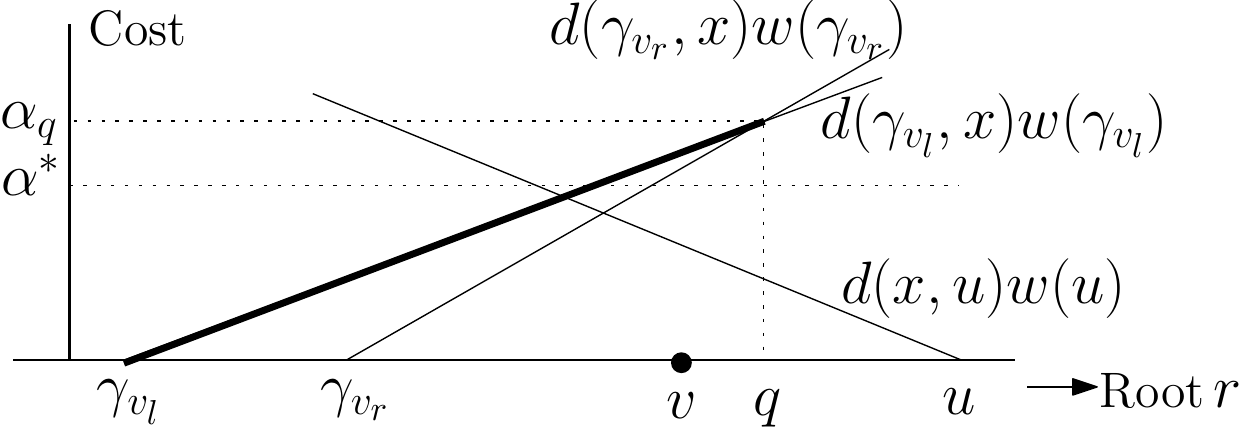}
%\hspace{2mm}
\includegraphics[height=2cm]{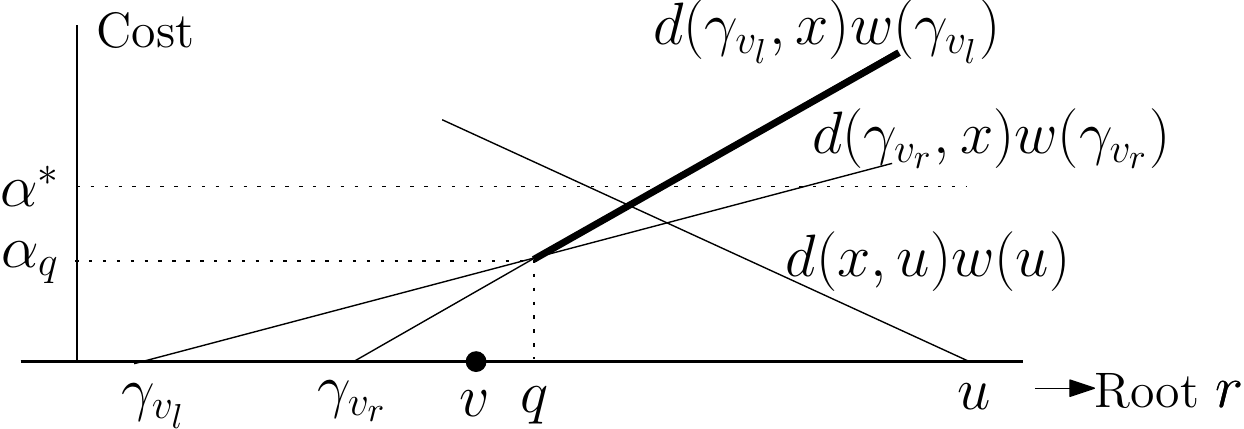}
\caption{The cost lines of $\gamma_{v_l} \in B(v_l)$ and $\gamma_{v_r} \in B(v_r)$ intersect at $q$
above $v$:
(Left) Cost $\alpha_q$ at intersection $q$ is higher than $\alpha^*$ ($\alpha_q>\alpha^*$); 
(Right) $\alpha_q<\alpha^*$.
}
\label{fig:++c}
\end{figure}
To this end,
we first find the intersection $q =c'(\gamma_{v_l}, \gamma_{v_r}) \in \pi[v,r]$ of the two cost lines
$d(\gamma_{v_l},x)w(\gamma_{v_l})$ and $d(\gamma_{v_r},x)w(\gamma_{v_r})$,
and its cost $\alpha_q= \alpha'(\gamma_{v_l}, \gamma_{v_r})$,
assuming the condition for
(\ref{eqn:equalcost2}) is met.
We then test $\alpha_q$-feasibility.
If $\alpha_q \geq \alpha^*$, as in the left figure of Fig.~\ref{fig:++c},
then we set $\gamma'_{v} =\gamma_{v_l}$ (resp. $\gamma_{v}=\gamma_{v_r}$)
if $w(\gamma_{v_l}) \leq w(\gamma_{v_r})$ (resp. $w(\gamma_{v_l}) > w(\gamma_{v_r})$).
If $\alpha_q < \alpha^*$, on the other hand, as in the right figure of Fig.~\ref{fig:++c},
then we set $\gamma'_{v}=\gamma_{v_r}$ (resp. $\gamma_{v}=\gamma_{v_l}$)
if $w(\gamma_{v_l}) \leq w(\gamma_{v_r})$ (resp. $w(\gamma_{v_r}) <w(\gamma_{v_l})$).
In order to find the true critical vertex $\gamma_{v}$ in place of $\gamma'_{v}$,
we need to take $v$ into consideration as well.
This time we use $\alpha'(v_a, v)$ of (\ref{eqn:equalcost4}) instead of (\ref{eqn:equalcost2}).
In the future we will be testing vertices $u \notin T(v)$ to see if the cost of the intersection 
between $d(x,u)w(u)$ and $d(\gamma_{v},x)w(\gamma_{v})$ 
is lower than $\alpha^*$ or not.
We must choose the critical vertex that gives the highest cost near $\alpha^*$,
which is indicated by a thick line segment in Fig.~\ref{fig:++c}.

In any case, we need to perform a constant number of feasibility tests per vertex visited.
Whenever an $\alpha$-feasibility test in (\ref{eqn:test10})
succeeds (resp. fails),
we update $\overline{\alpha}$ (resp. $\underline{\alpha}$) to $\alpha$.
\begin{lemma} \label{lem:finalAlpha}
The optimal cost $\alpha^*$ equals $\overline{\alpha}$ at the end of the above steps.
\end{lemma}
\begin{proof}
It was shown by Kariv and Hakimi~\cite{kariv1979b} that $\alpha^*$ has the value
$d(u,v)/(1/w(u)+1/w(v))$ for some pair of vertices $u$ and $v$.
See Lemma~\ref{lem:eqCost}(a). 
It is clear that the above steps partition the vertex set to $p$ maximal subsets
$\{V_i\mid i=1,2,\ldots, p\}$
such that each subset $V_i$ can be $\overline{\alpha}$-covered by a center.
The value of $\overline{\alpha}$ at the end of our algorithm is
from the last ${\alpha}$-feasibility test that reduced $\overline{\alpha}$.
Assume that $\alpha^*< \overline{\alpha}$ and there is a pair of
vertices $u$ and $v$ in the same subset such that  $\alpha^*= d(u,v)/(1/w(u)+1/w(v))$,
but we haven't tested them. 
We derive a contradiction to this assumption.

Let us examine how each $V_i$ was formed.
Bottom-up, we constructed the upper bound on the cost functions of the vertices.
As we moved higher, we tested (\ref{eqn:test10}) %and (\ref{eqn:test11})
for each vertex $v$ in $V_i$,
and $\overline{\alpha}$ was updated as a result.
Since critical vertices $\gamma_{v_l}$ (resp. $\gamma_{v_r}$) is used in
(\ref{eqn:test10}), %(resp. (\ref{eqn:test11})),
this $\overline{\alpha}$ is the smallest possible cost to cover $V_i$.
 \QED
\end{proof}
%This completes the proof of Theorem~\ref{thm:main}.

%%%%%%
\subsection{General tree networks}\label{sec:general}
We use spine tree decomposition (STD),
reviewed in Sec.~\ref{sec:std},
 for general (unbalanced) tree networks.
The counterparts to Theorems~\ref{thm:feasibility} and \ref{thm:practical} hold
with the same complexities.
\begin{theorem}\label{thm:general1}
\begin{enumerate}
\item[(a)]
We can test $\alpha$-feasibility in $O(p\log (n/p))$ time,
excluding the preprocessing,
which takes $O(n \log n)$ time.
\item[(b)]
WD$p$C for the general tree networks with $n$ vertices can be solved in 
 $O(n \log n+p^2\log^2 (n/p))$ time.
\end{enumerate}
\end{theorem}
\begin{proof}
Part(a) can be proved in essentially the same way
as we proved Theorem~\ref{thm:feasibility} in Sec.~\ref{sec:test}.
Instead of working directly on the given tree $T$,
we first construct ${\it STD}(T)$ and compute upper envelopes at its nodes.
The concepts of the $\ominus$-subtree and $\oplus$-subtree can be carried
over to ${\it STD}(T)$.
One complication is that we need to work on a group of $\ominus$-branches,
instead of single $\ominus$-subtrees,
but we can process them in the same order of time as in the balanced tree case.
Part (b) is implied by part (a) by the main theorem in Megiddo~\cite{megiddo1979}.
\QED
\end{proof}

As for the counterpart to Theorem~\ref{thm:main}, 
we need to use AKS-like sorting networks~\cite{ajtai1983,goodrich2014,paterson1990,seiferas2009},
as in~\cite{cole1987}.

\begin{theorem}\label{thm:general2}
WD$p$C for the general tree networks with $n$ vertices can be solved in
 $O(n \log n+p\log^2 n\log (n/p))$ time.
\end{theorem}
\begin{proof}
~[Informal]
Let us first analyze how many times we need to perform feasibility tests
when {\it STD}$(T)$ is used for a non-balanced tree network.
Let $n_l$ be the number of vertices in the spines at level $l$,
so that we have $\sum_{l=1}^{\lambda} n_l=n$,
where $\lambda$ is the number of levels in {\it STD}$(T)$.
We now consider one particular spine $\sigma_l$ at level $l$.
Let $v_i$ and $v_k$ be two vertices on $\sigma_l$,
from which branches $B_i$ and $B_k$ hang.
Assume first that both $B_i$ and $B_k$ are $\ominus$-branches,
and let $\gamma_{v_i}$ (resp. $\gamma_{v_k}$) be the $\alpha^*$-critical vertices in
$B_i$ (resp. $B_k$).
If $\gamma_{v_i}$ is at distance $d_i$ from $v_i$,
then we map it onto $\sigma_l$ at distance $d_i$ from $v_i$.
\begin{figure}[ht]
\centering
\includegraphics[height=22mm]{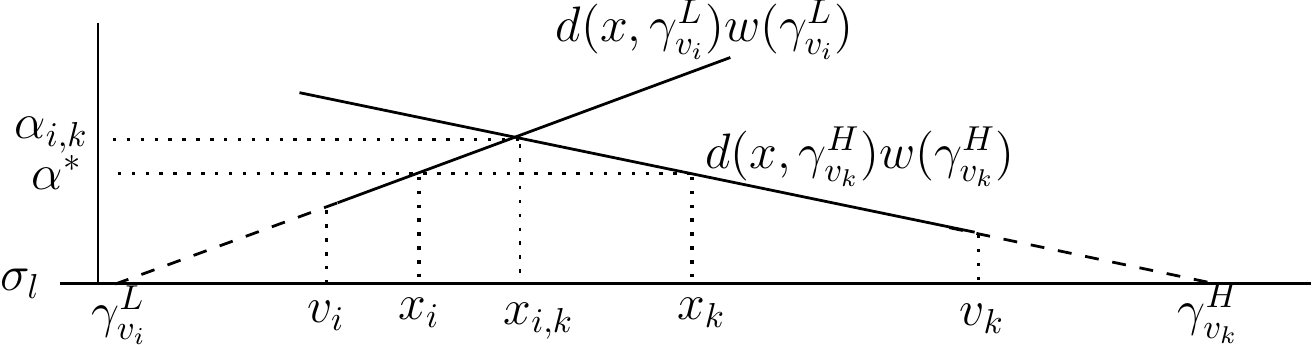}
\caption{$B_i$ and $B_k$ are each a $\ominus$-branch. 
}
\label{fig:--d}
\end{figure}
There can be up to two such positions on $\sigma_l$ (or its extension if it is not long enough),
and we call the lower (resp. higher)\footnote{Lower (resp. higher) means farther (resp. nearer) from/to
the root.}
one $\gamma^L_{v_i}$ (resp. $\gamma^H_{v_i}$).
Fig.~\ref{fig:--d} illustrates $\gamma^L_{v_i}$ and $\gamma^H_{v_k}$.
In this figure each cost function $d(x,\gamma^L_{v_i})w(\gamma^L_{v_i})$
is represented by a solid and a dashed line,
where the solid (resp. dashed) part shows its value on $\sigma_l$ (in $B_i$).
Similarly for the cost function $d(x,\gamma^H_{v_k})w(\gamma^H_{v_k})$.
In this figure, they meet at $x_{i,k}$ on $\sigma_l$,
and at this point the cost is $\alpha_{i,k} >\alpha^*$.
This implies that $x_i \prec x_k$,
where $x_i$ (resp. $x_k$) is the point on $\sigma_l$ where the cost of
$\gamma^L_{v_i}$ (resp. $\gamma^H_{v_k}$) is $\alpha^*$.
This in turn means that a single center cannot $\alpha^*$-cover both $\gamma^L_{v_i}$
and $\gamma^H_{v_k}$.
If we had $\alpha_{i,k} \leq \alpha^*$,
then a center would cover both of them.

Consider next the case where $B_i$ is a $\ominus$-branch and $B_j$ is a $\oplus$-branch,
as shown in Fig.~\ref{fig:+-d}.
\begin{figure}[ht]
\centering
\includegraphics[height=22mm]{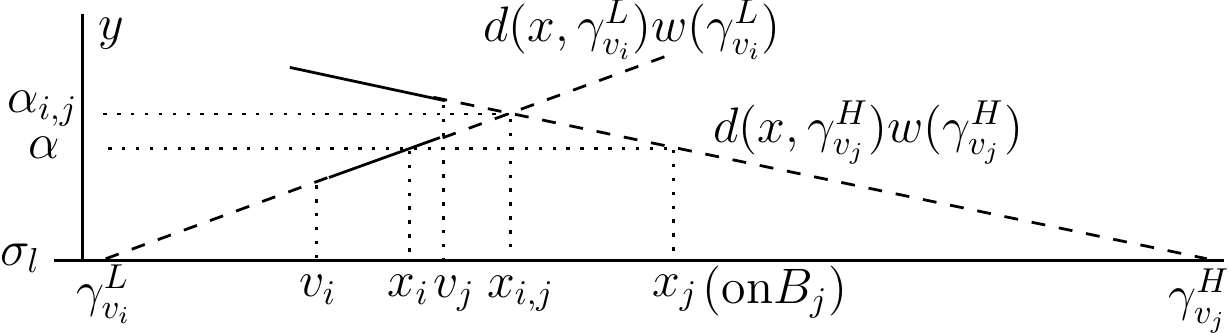}
\caption{
Point $x_j$ is the mapped image onto $\sigma_l$ of the highest center in $B_j$.
}
\label{fig:+-d}
\end{figure}
In this case, the dashed part of the cost function $d(x,\gamma^H_{v_j})w(\gamma^H_{v_j})$
takes the value $\alpha^*$ at $x_j\in B_j$,
which means that $B_j$ is a $\oplus$-branch.
The two cost functions $d(x,\gamma^L_{v_i})w(\gamma^L_{v_i})$ and 
$d(x,\gamma^H_{v_j})w(\gamma^H_{v_j})$ intersect at $x_{i,j}$ in their dashed parts,
which implies that they meet in $B_j$.
Since the corresponding cost $\alpha_{i,j}$ is larger than $\alpha^*$ in this figure,
a center at $x_j\in B_j$ cannot $\alpha^*$-cover $\gamma_{v_i}$.

The above discussion implies that whether the cost at the intersection of two cost lines
is higher or lower than $\alpha^*$,
which can be tested by a feasibility test, 
determines if an additional center needs to be introduced or not.
Each feasibility test determines the relative order of $x_i, x_j, x_k$, etc.,
for all vertices on spine $\sigma_l$.
This is tantamount to sorting $x_i, x_j, x_k$, etc.,
which we can do by a sorting network, such as the AKS sorter.
By examining the sorted sequence, 
and scanning $\sigma_l$ from its lower end,
we can determine the number of centers needed on $\sigma_l$.

Finally, we need to find the $\alpha^*$-critical vertex that represents the part of spine $\sigma_l$
not covered by the centers introduced so far,
or the center that could cover additional vertices in the next higher spine.
Namely,
spine $\sigma_l$ may become a $\ominus$-branch or a $\oplus$-branch {\em vis-\`{a}-vis}
the next higher spine.
If it becomes a $\ominus$-branch,
there may be several candidates for the $\alpha^*$-critical vertex.
The situation is somewhat to that depicted in the left figure in Fig.~\ref{fig:++c},
where $\gamma_{v_l}$ and $\gamma_{v_r}$ are the two candidates.
The $\alpha^*$-critical vertex is whichever candidate whose cost line reaches $\alpha^*$ first,
i.e., at the lowest position.

If $\sigma_l$ becomes a $\oplus$-branch in the next higher spine,
we want to find the $\alpha^*$-critical vertex in $\sigma_l$ that can cover the ``farthest'' vertex
in the next higher spine.
Therefore, among the candidate critical vertices we pick the one whose cost line reaches $\alpha^*$ last,
i.e., at the highest position..

Following Megiddo~\cite{megiddo1983c},
for each spine we employ an AKS sorting network.
The number of inputs to the AKS sorting networks employed at level $l$ is thus $2n_l$.
Each such AKS sorting network has $O(\log n_l)$ layers of comparators,
and their sorted outputs can be computed with $O(\log n_l)$ calls to a feasibility test
with Cole's speed up~\cite{cole1987}.
The total number of calls at all levels $l=1,2,\ldots, \lambda$ with Cole's speed up is thus
$O(\sum_{l=1}^{\lambda} \log n_l)$.
Since $\sum_{l=1}^{\lambda} n_l = O(n)$,
we have $\sum_{l=1}^{\lambda} \log n_l \leq \lambda \log (n/\lambda) = O(\log^2 n)$.
Since each feasibility test takes $O(p\log (n/p))$ by Theorem~\ref{thm:feasibility}
(extended to {\it STD}$(T)$),
the total time spent by the feasibility tests is $O(p\log^2 \log (n/p))$.
In addition, we need time to compute the median at each layer of the AKS networks,
which is $O(n_l)$ per layer and $O(n_l \log n_l)$ at level $l$.
Summing this for all levels, we get
$O(\sum_{l=1}^{\lambda} n_l \log n_l) = O(n\log n)$. 
\QED
\end{proof}

%%%%%%
\section{Conclusion and Discussion}\label{sec:conclusion}
We have presented an algorithm for the weighted discrete $p$-center problem
for tree networks with $n$ vertices,
which runs in $O(n \log n+p \log^2 n \log (n/p))$ time.
This improves upon the previously best $O(n\log^2 n)$ time algorithm~\cite{cole1987}.
The main contributors to this speed up are spine tree decomposition,
which enabled us to limit the tree height to $O(\log n)$,
and the root-centric location policy,
which made locating centers simple.
Fractional cascading helped to shave a factor of $O(\log n)$
off the time complexity in Theorem~\ref{thm:feasibility}.
The $O(n\log^2 n)$ time algorithm~\cite{cole1987} and ours both
make use of the AKS sorting network~\cite{ajtai1983},
which is impractically large. 
However, recently AKS-like sorting networks with orders of magnitude reduced sizes
have been discovered~\cite{goodrich2014,seiferas2009},
and further size reduction in the not-so-distant future may make the above algorithms more practical.
We also presented a practical $O(n\log n + p^2\log^2(n/p))$ time WD$p$C algorithm,
which improves upon the $O(n\log^2 n\log\log n)$ time algorithm~\cite{megiddo1983c}
when $p=O(\sqrt{n})$.

In Lemma~\ref{lem:envelope} we showed that it takes $O(n\log n)$ time and space
to compute the set of bending point sequences for the upper envelopes at all the vertices.
Suppose that the weight of a vertex is increased arbitrarily,
which could influence the locations of some centers,
if the vertex becomes critical for a center.
We can test this situation without updating the upper envelopes,
and thus without increasing the time requirement.
Therefore, every $p$-center query with the weight of one vertex
arbitrarily increased can be answered in $O(p\log (n/p)\log n)$ time.
This result realizes a sub-quadratic algorithm for the minmax regret $p$-center problem
in tree networks~\cite{averbakh1997}.

\end{document}